# Comment on ''Photothermal radiometry parametric identifiability theory for reliable and unique nondestructive coating thickness and thermophysical measurements'' [J. Appl. Phys. 121(9), 095101 (2017)]

J.-C. Krapez[1a)], F. Rigollet[2]

[1]ONERA, The French Aerospace Lab, DOTA, Salon de Provence, France
[2]Aix Marseille Univ., CNRS, IUSTI, Marseille, France

A recent paper [X. Guo, A. Mandelis, J. Tolev and K. Tang, J. Appl. Phys., **121**, 095101 (2017)] intends to demonstrate that from the photothermal radiometry signal obtained on a coated opaque sample in 1D transfer, one should be able to identify separately the following three parameters of the coating: thermal diffusivity, thermal conductivity and thickness. In this comment, it is shown that the three parameters are correlated in the considered experimental arrangement, the identifiability criterion is in error and the thickness inferred therefrom is not trustable.

Guo et al.[1] recently presented what they consider "a detailed reliability analysis of estimated parameters to a three-layer theoretical model of photothermal radiometry frequency domain signals by applying parameter identifiability conditions". The model they used for representing the photothermal radiometry experiment and the dynamic temperature of the material front surface is a one-dimensional four-layer model: the air layer and the substrate layer are semi-infinite and a coating of thickness $L_2$, diffusivity $\alpha_2$ and conductivity $\kappa_2$ is associated with a thin "roughness layer"; all three solid layers are considered opaque. The question the authors want to address is the multiparameter identifiability depending on whether the coatings parameters are considered separately { $L_2$, $\alpha_2$, $\kappa_2$ } or grouped according to { $Q_2$, $P_2$ } with $Q_2 = L_2/\sqrt{\alpha_2}$ and $P_2 = \kappa_2/\sqrt{\alpha_2}$. One can recognize in $Q_2$ the square root of thermal transit-time (or diffusion-time) through the coating and in $P_2$ the effusivity of the coating. A third possibility the authors considered is the two-parameter sequence $\{Q_2, b_{32}\}$ where $b_{32}$ is the ratio between the effusivity of the substrate and the coating ($b_{32} = P_3/P_2$). Based on a sensitivity analysis of the amplitude and phase of the photothermal signal to the $L_2$, $\alpha_2$, $\kappa_2$, the authors pretend that this three-parameter set may be identifiable, depending on the frequency set choice. This is actually *not the case* as it will be shown below.

A significant number of papers have already addressed the problem of identifiability regarding the parameters of a coating in the course of a photothermal measurement, with periodic excitation[2-5], pulsed excitation[6-12] or both[13]. Of benefit are also some works on the thermal characterization of a given layer in a multilayer stacking[14-16].

The authors appropriately mentioned that the expression of the radiometric signal in eq. (1)-(3) in Ref. 1 could be transformed in such a way that the coating properties intervene through a set of *two parameters only*: $Q_2$ and $P_2$, as it appears in eq. (6)-(7) in Ref. 1. Applying the quadrupole formalism would have highlighted this feature much easier (see Ref. 17, 18 and applications[2, 5-11,13-16]). As a matter of fact, when deprived from internal sources, the layer contribution is fully represented by the quadrupole[17,18]:

$$M = \begin{bmatrix} \cosh Q_2\sqrt{p} & (P_2\sqrt{p})^{-1}\sinh Q_2\sqrt{p} \\ P_2\sqrt{p}\sinh Q_2\sqrt{p} & \cosh Q_2\sqrt{p} \end{bmatrix} \quad (1)$$

where $p$ represents either the Laplace variable (time domain analysis) or $2i\pi f$ (frequency domain analysis). Although there is no one-to-one relation between time values and frequency values, many properties observed in one domain can be transposed into the other one (which motivates cross-fertilization of both research fields). The concept of *absolute non-identifiability conditions* is an example. This means that if we are facing a combination of parameters such that non-identifiability is observed in the Fourier domain *for any set of frequency values*, then non-identifiability is observed in the time domain *for any set of time values*, and vice-versa. This is exactly what is encountered with the combination of parameters { $L_2$, $\alpha_2$, $\kappa_2$ } because they are *fully correlated*.

The conditions of identifiability of $Q_2$ and $P_2$ (or related parameters) were analysed in many instances[2-13] (in particular, $Q_2$ cannot be identified when the coating and the substrate have the same effusivity; the interface between them remains hidden in any photothermal measurement, let it be pulsed or modulated). The thickness $L_2$ only appears in $Q_2$, bound to diffusivity $\alpha_2$ (see eq. (1)). It means that once $Q_2$ is identified, $L_2$ can be evaluated only if $\alpha_2$ is known from elsewhere, or symmetrically, $\alpha_2$ can be evaluated only if $L_2$ is known (what was done in ref. [3, 4], among others). Identifying $P_2$ in addition to $Q_2$ does not help much since, by multiplying or dividing the latter two, one gets either the heat capacitance of the coating $\rho_2 C_2 L_2$ or its thermal resistance $L_2/\kappa_2$. Again, for getting the value of $L_2$ one needs to know from elsewhere either the volumetric heat capacity or the conductivity of the coating (indeed, knowing the effusivity is not enough). Actually, one can *by no means*

---

[a)] Email: krapez@onera.fr

combine $Q_2$ and $P_2$ so as to get separately the three parameters $L_2$, $\alpha_2$ and $\kappa_2$. In conclusion of the consequences of an initial identification of the parameter set $\{Q_2, P_2\}$, $L_2$ always appears bound to either $\alpha_2$, $\rho_2 C_2$ or $\kappa_2$, without any possibility of evaluating it except through an *independent measurement* of *either* $\alpha_2$, $\rho_2 C_2$ or $\kappa_2$. Let us note that all these comments are actually valid for any layer of finite thickness in a multilayer component. Performing a 2D photothermal experiment involving lateral diffusion, not only through-thickness diffusion, or an experiment involving internal sources inside the layer under study would provide an additional information that could help identifying separately $L_2$, $\alpha_2$ and $\kappa_2$, but this was not the subject of the paper under discussion.

Can it happen that with a good choice of frequencies the three-parameter set $\{L_2, \alpha_2, \kappa_2\}$ can be evaluated directly? This was the main topic of ref. 1 and *the answer is no*. The fact that the amplitude (or the phase) of the signal is totally defined (regarding the coating properties) by the two-parameter set $Q_2 = L_2/\sqrt{\alpha_2}$ and $P_2 = \kappa_2/\sqrt{\alpha_2}$ has the consequence that the sensitivity vectors to $L_2$, $\alpha_2$ and $\kappa_2$ are *linearly dependent*. As a matter of fact, since $G = G(Q_2, P_2)$, where $G$ is for amplitude or phase, the partial derivatives with respect to $L_2$, $\alpha_2$, $\kappa_2$ can be obtained, according to the chain rule, as:

$$\frac{\partial G(f_i)}{\partial \alpha_2} = -\frac{L_2}{2\alpha_2^{3/2}} \frac{\partial G(f_i)}{\partial Q_2} - \frac{\kappa_2}{2\alpha_2^{3/2}} \frac{\partial G(f_i)}{\partial P_2}$$
$$\frac{\partial G(f_i)}{\partial \kappa_2} = \frac{1}{\alpha_2^{1/2}} \frac{\partial G(f_i)}{\partial P_2} \qquad , \forall f_i \qquad (2)$$
$$\frac{\partial G(f_i)}{\partial L_2} = \frac{1}{\alpha_2^{1/2}} \frac{\partial G(f_i)}{\partial Q_2}$$

The following *linear dependence* regarding the three sensitivity vectors is readily inferred:

$$2\alpha_2 \frac{\partial G(f_i)}{\partial \alpha_2} + \kappa_2 \frac{\partial G(f_i)}{\partial \kappa_2} + L_2 \frac{\partial G(f_i)}{\partial L_2} = 0 \quad , \forall f_i \qquad (3)$$

We stress the fact that this relation is valid *for any frequency set* and *for any combination of $L_2$, $\alpha_2$ and $\kappa_2$ values* (despite the non-linear influence of these parameters on temperature). $L_2$, $\alpha_2$ and $\kappa_2$ are thus *fully correlated* and consequently *not identifiable*. By the way, the linear dependence in eq. (3) is easily observed on the sensitivity curves plotted in fig. (5)-(6), whatever the frequency and thickness values. This corresponds to one of the pathological cases described by one of the authors in Ref. 4:"the best way to make an inversion problem effective is to use only experimental data in the regions where the sensitivity coefficients are high, not proportional, nor almost proportional, nor, in general, constitute a linear combination".

The consequence of eq. (3) is that the determinants in eq. (17)-(18) in Ref. 1 are *exactly zero* whatever the *triplet* of frequency values $f_i$ and *the combination of $L_2$, $\alpha_2$ and $\kappa_2$ values*. It is unclear on how the authors did build the identifiability maps in fig. 2, 3. These maps are intended to show combinations of parameters and frequencies that would (allegedly) allow a joint identification of $L_2$, $\alpha_2$, $\kappa_2$. "The black diamonds in the figure indicate the locations where the identifiability condition is not met, e.g., the three parameters are linearly dependent (zero determinant of the sensitivity coefficients matrix)". However these determinants involve *three* frequency values whereas each diamond in the maps is related to only *one* frequency value. Anyway, what eq. (3) shows is that any "identifiability map" should be fully covered with diamonds, highlighting that the identifiability condition is *never met*. If the authors found in some circumstances that the determinants in eq. (17)-(18) in Ref. 1 are not strictly zero, is most certainly due to round-off errors. As a matter of fact the sensitivity coefficients $\partial G(f_i)/\partial \beta$ were computed numerically by a finite difference approximation of the true derivatives. The consecutive round-off error in each *numerically approached* sensitivity coefficient can cause the right member of eq (3) not be equal *strictly* to zero, and let the authors think that there is no linear dependence. If the authors computed the equation (3) for *any* frequency, they would find without a doubt a right member *very close to zero*, which is a warning that a linear dependence may be present.

Another (hopeless) idea would be to identify the parameter set $\{L_2, \alpha_2, P_2\}$. Here, the sensitivity vectors to diffusivity $\alpha_2$ and to thickness $L_2$ are *linearly dependent*:

$$2\alpha_2 \frac{\partial G(f_i)}{\partial \alpha_2} + L_2 \frac{\partial G(f_i)}{\partial L_2} = 0 \quad , \forall f_i \qquad (4)$$

causing the corresponding determinants to be *zero as well*. $L_2$, $\alpha_2$, $P_2$ are *fully correlated* and *not identifiable* either.

Finally, the authors present an allegedly experimental validation of $\{L_2, \alpha_2, \kappa_2\}$ identifiability (§ B.2) (remark: the first layer was thermally thin, thus allowing to merge it with the "actual" coating thickness $L_2$ in $L = L_1 + L_2$). How the authors have done to identify the total coating thickness $L$ separately from $\alpha_2$ and $\kappa_2$ remains a mystery. These estimations are indeed theoretically and without a doubt completely impossible, whatever the methodology. In the least squares approach used by the authors, the information matrix $S^T S$ built with the three sensitivity vectors (let it be with only three frequencies or many more), needs to be inverted. The linear dependence between them (eq (3)), makes the determinant of $S^T S$ equal to zero, preventing its inversion. The only explanation for the (theoretically impossible) estimation is again in the residual round-off errors whose consequence is to hide the parameter correlation. Nonetheless, this parameter correlation has strong consequences on the parameter identification and on the uncertainty estimation, even though the *numerically computed* $S^T S$ is *not strictly singular* (in such a case, during the estimation process, the user generally receives a 'bad conditioning" message, which should warn him of an identification failure). Unavoidably, the retrieved parameter uncertainties are very high. Yet, no explanation was given about the determination of the uncertainty range for $L$ (neither about the bias due to the supposed known – fixed – parameters, as for example the reflectance values when dealing with the signal amplitude ratio) The authors should have given the diagonal elements of the covariance matrix (which are undoubtedly huge). Fundamentally, since the coating diffusivity and its thickness *are not separable, any couple of* $L_2$ *and* $\alpha_2$ *values* leading to the value of the square root of thermal-diffusion-time $Q_2 = L_2/\sqrt{\alpha_2}$ that fits in the considered frequency range *would be acceptable*. That is to say any other value of $L$ than the one claimed in the paper would be admissible. In other words, the announced thickness values have actually an *infinite uncertainty range*. The authors' statement "This result is significant for confirming the uniqueness of the measurement" is then wrong. The demonstration provided here clearly emphasizes that the *identification is not unique* and hence not trustable.

On the other hand, to consider the determinant of the sensitivity matrix $S_A$ (or $S_P$) for establishing an identifiability criterion, as proposed by the authors, presents a serious limitation since, in the present case, it restricts the analysis to *three frequencies* as opposed to the many available frequencies. This is all the more surprising since, when considering experimental data in section III-B, the authors used 59 or 19 frequencies, not only 2 or 3 frequencies as mentioned in the sensitivity analysis section III-A. More powerful identifiability criteria were suggested in Ref. 19 and used e.g. in Ref. 5-16; one of them consists in maximizing the determinant of the matrix $S^T S$ or equivalently maximising the product of its eigenvalues. The determinant maximisation implies the minimization of the confidence region associated to the identified parameters. By the way, the diagonal of the covariance matrix $(S^T S)^{-1}$ provides the noise variance amplification factors for these parameters. Of interest is also the condition number of the matrix $S^T S$ [5,20]: getting huge values is a warning that something is wrong with the parameter choice.

We disagree with the discussion on the identifiability of the parameters $\{Q_2, P_2\}$ and $\{Q_2, b_{32}\}$. The authors relate *non-identifiability to the zero-crossing* of one or another curve of sensitivity vs. frequency. This is not a valuable criterion since the identification does not rely on a *single* frequency but many of them. Analysing the determinant of $S_A$ (or $S_P$) (which is based on *two frequencies* only - see eq. (19)-(22) in Ref. 1) is just a little better. Again, more interesting criteria are the condition number of the matrix $S_A^T S_A$ (or $S_P^T S_P$) or the diagonal of the covariance matrix. By the way, plotting the *reduced sensitivities* $\beta \partial G(f_i)/\partial \beta$, where $\beta$ is any parameter, instead of the absolute sensitivities $\partial G(f_i)/\partial \beta$ (as in fig. 10 for example), makes their comparison easier since they share the same units.

In conclusion, with the considered experimental arrangement, one can identify at best two "parameters" related to the coating: the effusivity ratio $b_{32}$ and, if the latter is not too close to 1, the square root of thermal transit-time $Q_2$ (the frequency range has to be optimized for minimizing their uncertainties). Next, either the coating thickness $L_2$ or its diffusivity has to be known for inferring the other one from $Q_2$ (see e.g. Ref. 3, 4). If the effusivity of the substrate is known, one can infer from $b_{32}$ the one of the coating, $P_2$. This offers the possibility of evaluating $L_2$ from $Q_2$ together with an independent knowledge of either conductivity or volumetric heat capacity (or the opposite). The possibility claimed in ref 1 that the coating thickness can be identified separately from the thermophysical properties (in particular the diffusivity) is false.